\documentclass[reprint,superscriptaddress]{revtex4-2}

\usepackage{soul}
\usepackage{graphicx}
\usepackage{dcolumn}
\usepackage{bm}
\usepackage{xcolor}
\usepackage{hyperref}
\usepackage{subfigure}
\usepackage[mathlines]{lineno}
\usepackage{amsmath}

\begin{document}

\preprint{APS/123-QED}

\title{Pushy Random Walk: A Minimal Model for Transport in Deformable Media}

\author{O. Lauber Bonomo}
\email[Corresponding author: ]{o.lauber@nyu.edu}

\affiliation{Center for Urban Science and Progress, Tandon School of Engineering, New York University, Brooklyn, New York, USA}

\affiliation{School of Chemistry, Center for the Physics and Chemistry of Living Systems,
and the Sackler Center for Computational Molecular Materials Science, Tel Aviv University, 6997801 Tel Aviv, Israel}

\author{I. Shitrit}
\email{itamarshtrit@mail.tau.ac.il}

\affiliation{School of Chemistry, Center for the Physics and Chemistry of Living Systems,
and the Sackler Center for Computational Molecular Materials Science, Tel Aviv University, 6997801 Tel Aviv, Israel}

\author{S. Reuveni}
\email{shlomire@tauex.tau.ac.il}

\affiliation{School of Chemistry, Center for the Physics and Chemistry of Living Systems,
and the Sackler Center for Computational Molecular Materials Science, Tel Aviv University, 6997801 Tel Aviv, Israel}

\author{S. Redner}
\email{redner@santafe.edu}

\affiliation{Santa Fe Institute, 1399 Hyde Park Road, Santa Fe, New Mexico 87501, USA}

\date{\today}

\begin{abstract}
We introduce the pushy random walk, where a walker can push multiple obstacles, thereby penetrating large distances in environments with finite obstacle density. This process provides a minimal model for experimentally observed interactions of active particles with dense, deformable media. Using scaling arguments and numerical simulations, we show that in one dimension the walker carves out an obstacle-free cavity whose length grows subdiffusively with time. In two dimensions, increasing obstacle density drives a transition from free diffusion to localized behavior, where the walker is trapped within a cavity whose radius again grows subdiffusively with time. These results show how tracer-induced rearrangements qualitatively reshape transport in crowded media.
\end{abstract}


\maketitle

The “ant in a labyrinth”~\cite{deGennes1976} is a colloquial term for a random walk that moves in a medium with a fixed density of obstacles.  If their density is low, the mean-square displacement of the walk grows linearly with time, but with a diminished amplitude as the obstacle density increases.  When the obstacle density is high, the walk is confined to a finite domain. When the accessible region for the walk is at the percolation threshold, anomalous behavior arises in which the mean-square displacement grows sublinearly with time. This unusual behavior sparked considerable interest in percolation-controlled anomalous diffusion (see, e.g., \cite{deGennes1976,AlexanderOrbach1982,GefenAharonyAlexander1983,HavlinNossal1984,Kesten1986,HavlinBenAvraham1987,GrimmettKestenZhang1993,stauffer2018introduction,Barlow2004}). 

Models of this genre assume the environment contains immobile obstacles. Recent experiments by Altshuler \textit{et al.}~\cite{Altshuler2024} reveal qualitatively new behavior arises when obstacles can be displaced by collisions with a random walker. As a first step toward understanding this coupling, Bonomo \textit{et al.}~\cite{Lauber2023,Lauber2024} introduced the \emph{Sokoban} random walk, in which the walk can displace at most single obstacles. This coupling has a strong dynamical consequence: in one, two, and three dimensions the Sokoban walk inevitably becomes trapped after a finite displacement~\cite{Lauber2023,singh2026sokoban,shitrit2026pushing}, with the typical travel distance decreasing as the obstacle density increases~\cite{Lauber2023,shitrit2026pushing}. 

In contrast, in the experiments of Ref.~\cite{Altshuler2024}, the tracer can push \emph{clusters} of obstacles.  The effective resistance to tracer motion increases with cluster size, making larger rearrangements progressively harder to achieve. A basic question arises: what is the tracer dynamics in such a medium?  The goal of this Letter is to determine some intriguing features of this dynamics through the pushy random walk model that we introduce below.

\begin{figure}[t!]
\includegraphics[width=0.48\textwidth]{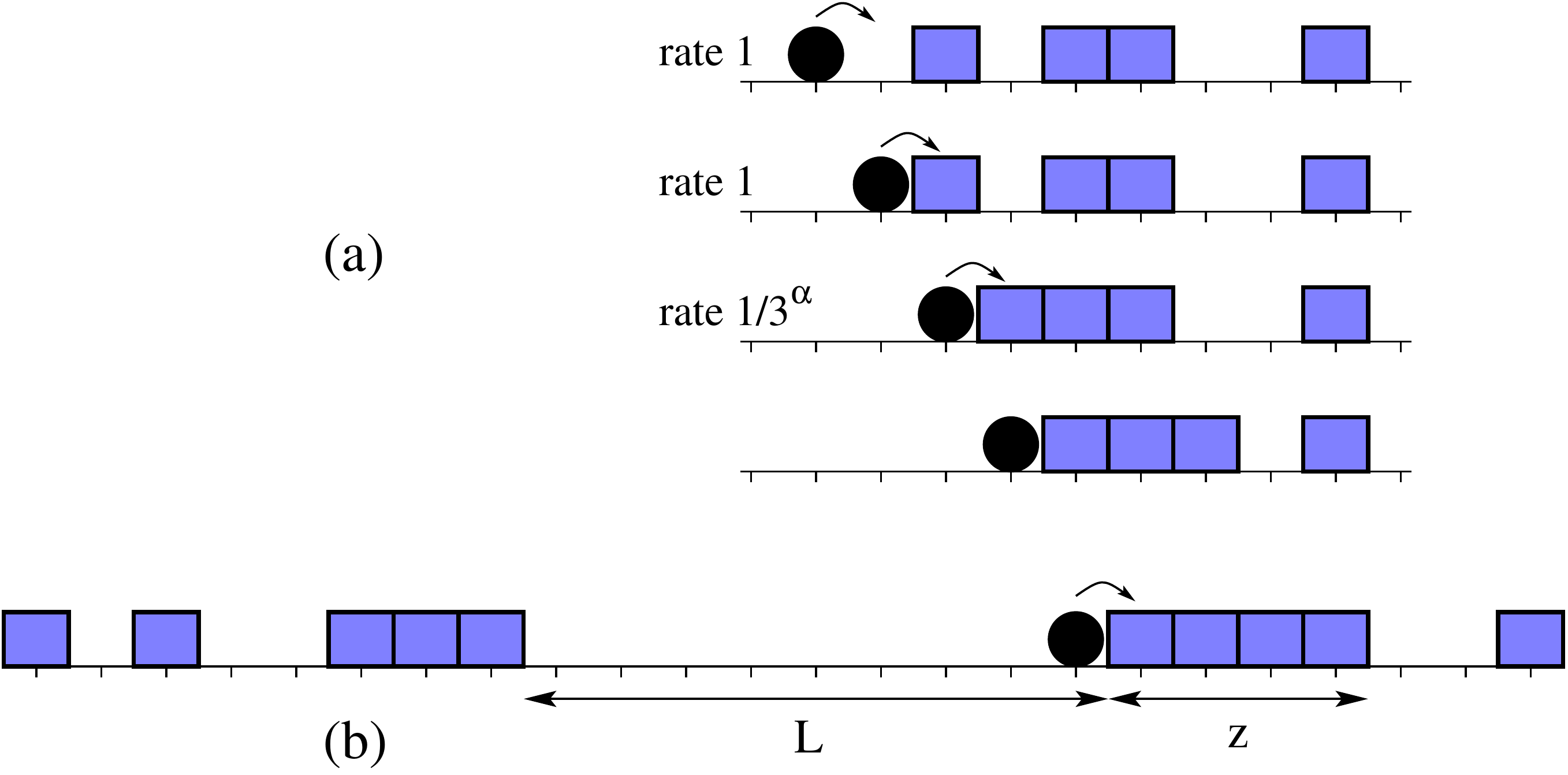}
\caption{(a) Illustration of a pushy random walk (circle) in one dimension.  An isolated unit-mass obstacle (square) is pushed one lattice spacing by the walk at rate 1. When a composite obstacle of mass 3 is created, it moves with rate $1/3^\alpha$ when hit by a pushy random walk. (b) A cavity of length $L$ and a crust of thickness $z=4$ on one side.}
\label{fig: 1D Model Illustration}
\end{figure}

\smallskip\noindent\textit{The model—} A random walk moves in a medium that contains unit-mass obstacles at density $\rho$.  When the walk hits an obstacle, both the walk and the obstacle may move one lattice spacing $a$ [Fig.~\ref{fig: 1D Model Illustration}].  By this pushing, an obstacle may merge with a more distant one to form a composite obstacle whose mass is the sum of the two initial masses. When a random walk hits a composite obstacle of mass $M$, the walk and the obstacle move by a distance $a$ with probability proportional to $M^{-\alpha}$. For example, in the third line of Fig.~\ref{fig: 1D Model Illustration}, the walk hops to the left with probability $\frac{1}{2}$.  If the walk attempts to hop to the right, also with probability $\frac{1}{2}$, this move is successful with probability $(1/3)^{\alpha}$, so that the walk and the obstacle move one lattice spacing to the right with total probability $\frac{1}{2}(1/3)^{\alpha}$.  With the complementary probability $\frac{1}{2}[1-(1/3)^{\alpha}]$, the walk and the obstacle do not move. 

In this picture, $\alpha$ quantifies the resistance of the medium to collective rearrangements: when the tracer encounters a block of mass $M$, the induced displacement is assumed to scale on average as $M^{-\alpha}$. We focus on the case $\alpha=1$, but our results readily generalize to arbitrary $\alpha\geq 0$.

\begin{figure}[t]
 \begin{center} 
 \includegraphics[width=0.3\textwidth]{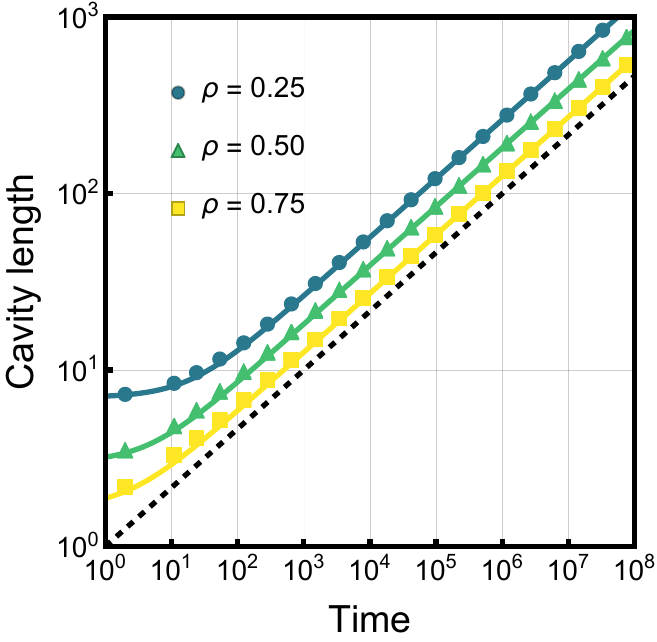}
  \caption{Cavity length versus time for a pushy random walk in 1D for different obstacle densities $\rho$. Solid lines give the theoretical prediction from Eq.~\eqref{L-1d sol} and symbols come from simulations averaged over $10^3$ walk trajectories. The dashed line shows the asymptotic behavior of $t^{1/3}$.}
\label{fig:1d_Simple_Clearing}
\end{center}
\end{figure}


\smallskip\noindent\textit{One dimension—} In one dimension, a pushy random walk gradually clears out a cavity of length $L(t)$. On either side of this cavity a solid “crust” of obstacles of thickness $z$ forms.  We determine its thickness by the criterion that the initial number of particles in a domain of length $L+2z$ equals the number of particles in a solid crust of thickness $2z$  (Fig.~\ref{fig: 1D Model Illustration}(b)).  This gives $(2z+L)\rho=2z$, or $z=\rho L/[2(1-\rho)]$.  If the random walk is adjacent to a crust of thickness $z$ and attempts to push on it, the crust moves one lattice spacing $a$ with probability $a/z$. In the long-time limit, the average time for the random walk to return to the crust on either side of the cavity, having started adjacent to it, is $a(L-a)/2D\approx aL/2D$~\cite{Redner2001}, where the diffusion coefficient is $D=1/2$ for a symmetric nearest-neighbor random walk with unit jump rate.

The  growth of the cavity is therefore accounted for by the rate equation
\begin{align}
\label{RE-1d}
  \frac{\Delta L}{\Delta t} \sim \frac{a\times 2a(1-\rho)/\rho L}{aL/2D} = \frac{4 D a(1-\rho)}{\rho L^{2}}\,,
 \end{align}
whose solution gives the subdiffusive growth 
\begin{align}
\label{L-1d sol}
L(t) = \big\{L(0)^3 + 12 Da\left[(1-\rho)/\rho\right]\,t\big\}^{1/3}\;,
\end{align}
with $L(0) = 1 + 2(1-\rho)/\rho$, as corroborated in Fig.~\ref{fig:1d_Simple_Clearing}. For general $\alpha$, $L(t)$ asymptotically grows as $t^{1/(2+\alpha)}$. When $\alpha\to 0$, the random walk has infinite “pushiness” and its motion reverts to pure diffusion.

\begin{figure}[t]
 \begin{center} 
 \includegraphics[width=0.35\textwidth]{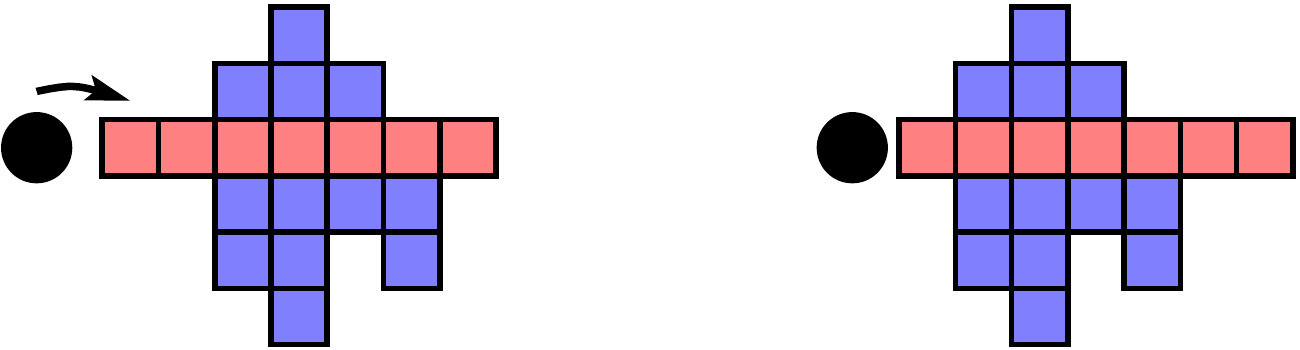}
  \caption{The result after a random walk in 2D pushes against a composite obstacle. Only the horizontal row of seven obstacles moves one lattice spacing.}
\label{fig:2d-motion}
\end{center}
\end{figure}

\smallskip\noindent\textit{Two and higher dimensions—} 
We generalize the pushing rule in higher dimensions as follows: We treat obstacles as frictionless blocks.  If a particle attempts to push a cluster of obstacles, only those obstacles that are collinear with the motion of the walk move by one lattice spacing. In the illustration of Fig.~\ref{fig:2d-motion}, the probability for this event is $\frac{1}{4}(1/7)^{\alpha}$; the factor $\frac{1}{4}$ gives the probability that the walk attempts to hop to the right on the square lattice, and the second factor of $(1/7)^\alpha$ gives the probability that the hopping attempt actually pushes the column of obstacles by one lattice spacing.
This rule should be viewed as a minimal coarse-grained extension of the one-dimensional dynamics. A description that includes transverse drag or shear would require a more detailed microscopic model for stress transmission and lateral rearrangements of neighboring obstacles, and is left for future work.

\begin{figure}[ht]
  \centerline{\includegraphics*[width=0.3\textwidth]{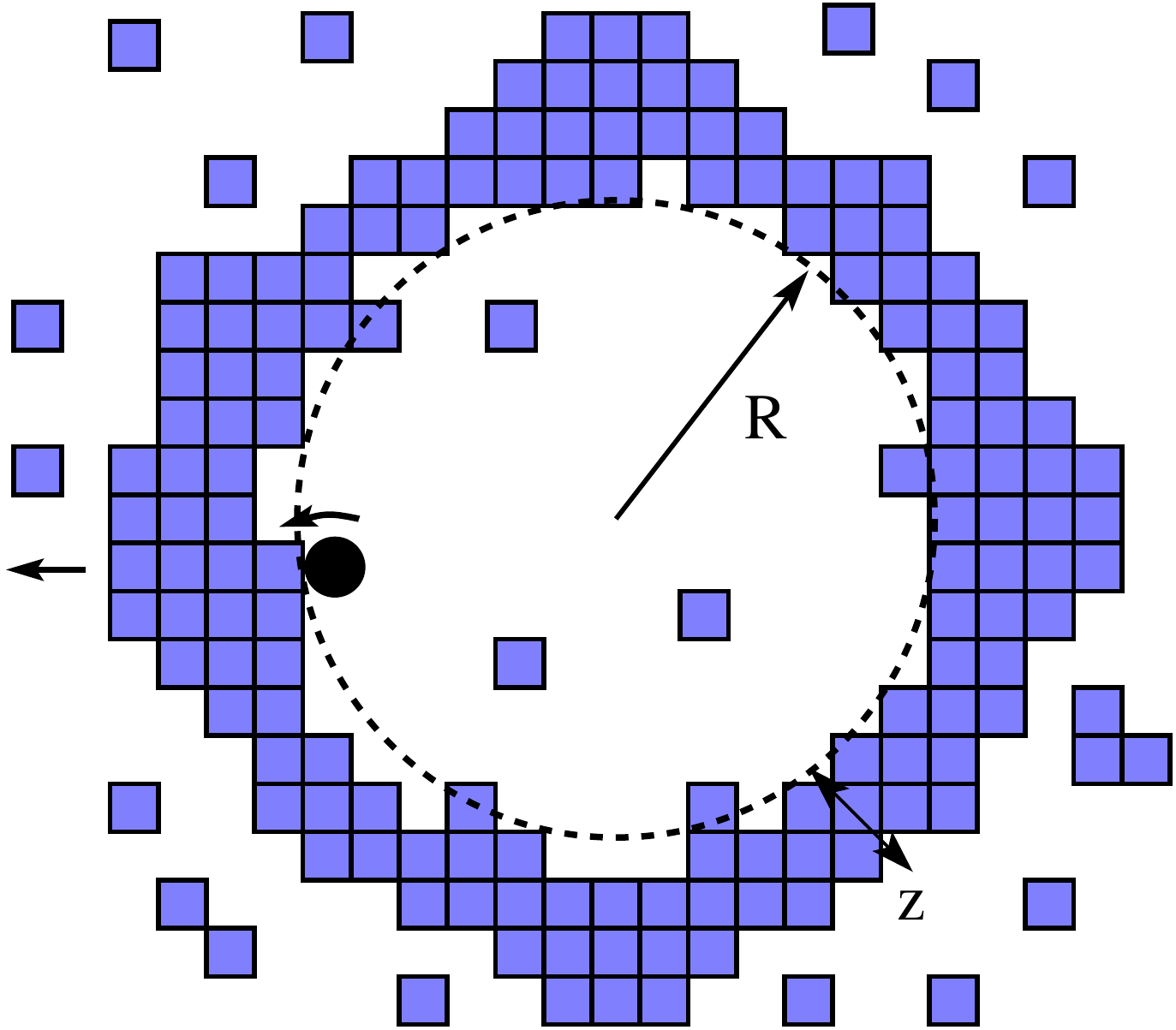}}
  \caption{A cavity in two dimensions with a surrounding crust. The random walk is about to push an obstacle of four contiguous particles horizontally to the left by one lattice spacing.}
\label{fig:2d}
\end{figure}

Since a random walk is recurrent in two dimensions~\cite{Feller1968}, it seems plausible that it will eventually clear out (perhaps incompletely) a compact cavity that is bounded by an annular crust of obstacles. Using this picture, we adapt the argument for one dimension to the two-dimensional case. We therefore suppose that the walk carves out a roughly circular cavity of radius $R$ that is surrounded by a solid annular crust of thickness $z$ (Fig.~\ref{fig:2d}). Applying the same argument as in one dimension, all particles within a circle of radius $(R+z)$ comprise a crust of thickness $z$ that surrounds a cavity of radius $R$.  This criterion gives $\pi \rho(R+z)^2= \pi[(R+z)^2-R^2]$, from which the crust thickness is 
\begin{align}
    \label{eq:z2d}
z=R[1/\sqrt{1-\rho}-1]\equiv A_2 R\,.
\end{align}
Crucially, the crust thickness is proportional to the cavity radius, as in the case of one dimension.

When the walk pushes against a point on the crust whose thickness is $z$, a single row of the crust will be displaced radially by a unit distance with probability $a/z$. Since there are $2\pi R/a$ possible locations for an obstacle along the crust, where $a$ is the obstacle radius, the motion of a single row of obstacles by one lattice spacing corresponds to the radius of the cavity increasing by $a/(2\pi R/a)$, on average.

\begin{figure}[t!]
  \centerline{\includegraphics[width=0.3\textwidth]{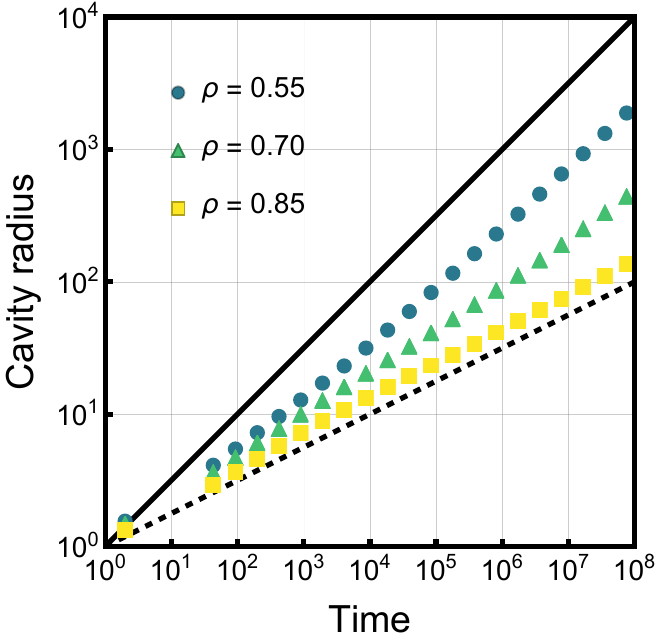}}
  \caption{Cavity radius $R(t)$ vs.\ time, for the pushy random walk on the 2D square lattice with different obstacle densities. Here, $R(t)$ is estimated as the square root of the number of distinct visited sites. The dashed and solid lines indicate $t^{1/4}$ and $t^{1/2}$ asymptotics, respectively. Symbols represent simulation data based on $2560$ walk trajectories.}
\label{fig:L2d}
\end{figure}

The other ingredient we need to write a rate equation for the cavity radius is the return time of a random walk that starts at the cavity boundary and returns to a point on this boundary.  For diffusion, we find this return time by solving the Poisson equation $D\nabla^2t =-1$, subject to the boundary conditions $t(r\!=\!R)=0$ and $\nabla t(r\!=\!0)=0$~\cite{Redner2001}. The solution is $t(r) = (R^2-r^2)/4D$. We use this result to approximate the return time of a random walk that starts at the cavity boundary, by taking its radial coordinate to be $r=R-a$. The return time to the boundary is asymptotically $aR/2D$. The generalization of the rate equation \eqref{RE-1d} to two dimensions therefore is
\begin{align}
\frac{\Delta R}{\Delta t } \sim \frac{(a^2/2\pi R) \times (a/A_2 R)}{a R/2D}
  = \frac{a^2 D}{\pi A_2 R^3}\,,
\end{align}
with solution 
\begin{align}
  R(t)=[R(0)^{4} + 4 a^2 D t/\pi A_2]^{1/4}\,.
  \label{2dSol}
\end{align}

For pushiness exponent $\alpha\ne 1$, Eq.~\eqref{2dSol} generalizes to $R(t)\sim t^{1/(3+\alpha)}$.  Curiously, for infinite pushiness, that is, $\alpha\!=\!0$, $R(t)$ does not grow as $t^{1/2}$, but rather as $t^{1/3}$.  This putatively slower than diffusive growth is based on the assumption that the cavity remains compact throughout its growth, which cannot be true in the infinite pushy limit. This inconsistency suggests that a transition must exist between subdiffusive tracer motion, corresponding to compact cavity growth, and free diffusion at a critical obstacle density $\rho_c$. To check this expectation, Fig.~\ref{fig:L2d} shows the time dependence of the cavity radius. There is subdiffusive $t^{1/4}$ growth of the radius for high obstacle density, $\rho>\rho_c$, and a $t^{1/2}$ growth law, characteristic of free diffusion, for $\rho<\rho_c$.

\begin{figure}[ht]
  \centerline{
  \includegraphics[width=0.35\textwidth]{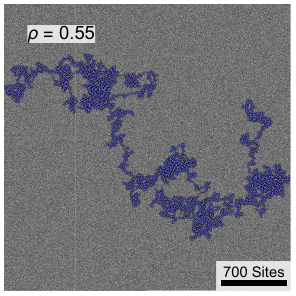}}
\centerline{  
\includegraphics[width=0.35\textwidth]{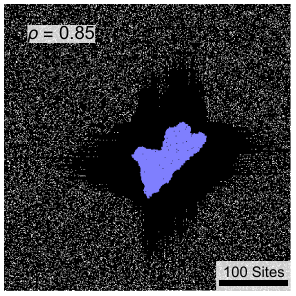}}
  \caption{Trajectories of the pushy random walk in the diffusive (top) and subdiffusive (bottom) regimes.  The apparently gray exterior area represents the unperturbed lattice with occupied sites black and unoccupied sites white.  Sites visited by the random walk are blue.  The solid black region in the lower panel is the crust. Further illustration of the pushy random walk is provided in Supplemental Material~\cite{SM}.}
\label{fig:transition-2d}
\end{figure}

The two representative trajectories of the pushy random walk in Fig.~\ref{fig:transition-2d} illustrate the change from diffusive exploration for $\rho<\rho_c$ to subdiffusive confinement within a compact cavity that is bounded by a solid crust. To understand how such a transition arises, we consider a random walk within such a cavity and determine the condition for the cavity to maintain its shape as it grows. 

\begin{figure}[t!]
  \centerline{\includegraphics[width=0.3 \textwidth]{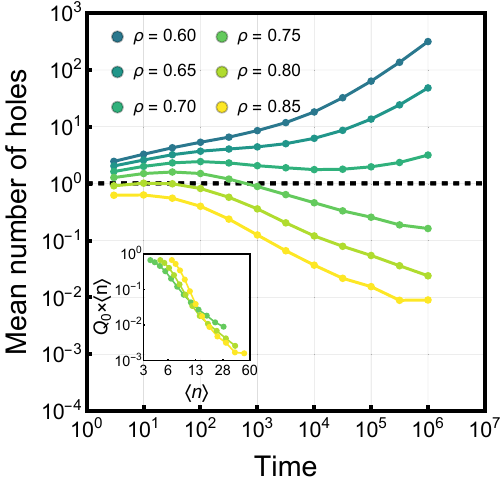}}
  \caption{Main: mean number of holes in the crust versus time for the pushy random walk on the 2D square lattice. Data are based on simulations of $8192$ trajectories. Inset: the product 
    $Q_0 \times \langle n\rangle$ versus $\langle n\rangle$.}
\label{fig:2d_Crust_Analysis}
\end{figure}

Since the crust grows stochastically, its thickness at any point along its circumference is a random variable whose mean value is $\langle n\rangle$. This average is determined by the condition that at a fixed angular position, the average number of obstacles that are stacked radially on a solid crust equals its thickness $z$ (see Eq.~\eqref{eq:z2d}) divided by the obstacle size:
\begin{align}
  \label{n}
  \langle  n\rangle =\frac{z}{a}=\frac{A_2 R}{a}\,.
\end{align}
The crust consists of roughly $\mathcal{N}=2\pi R/a$ independent locations around its circumference where $\langle n\rangle$ particles are stacked, on average.  However, the number of particles at each location is a random variable with a nontrivial dependence on time and obstacle density. We therefore do not assume a specific parametric form for the full crust-thickness distribution, and instead focus on the vacancy probability $Q_0(\langle n\rangle)$, the probability that a given angular position on the crust contains no particles. Clearly, $Q_0(\langle n\rangle)$ must decay with $\langle n\rangle$.  It turns out that this decay is faster than $1/\langle n\rangle$ (see the inset to Fig.~\ref{fig:2d_Crust_Analysis}), an important fact that we will exploit momentarily.

Since there are $\mathcal{N}$ independent points along the crust circumference, a simple but crude criterion for the crust to be impenetrable is that the mean number of holes in the crust is less than 1:
\begin{align}
\label{criterion}
\mathcal{N} Q_0(\langle n\rangle) < 1\,.
\end{align}
In Fig.~\ref{fig:2d_Crust_Analysis}, we show that this criterion indeed demarcates the diffusion-subdiffusion transition. For obstacle densities $\rho > \rho_c\approx 0.71$, the mean number of holes is close to 1 for early times but eventually decreases, which indicates that a solid enclosing crust is being formed. In contrast, for $\rho<\rho_c$ the mean number of holes in the crust is always greater than 1.  These holes allow the walk to escape the cavity and diffuse. 

From the criterion $\mathcal{N} Q_0(\langle n\rangle) < 1$, we can write a formal expression for the critical density $\rho_c$.  We first write Eq. \eqref{criterion} explicitly:
\begin{align*}
\frac{2 \pi R}{a}\;Q_0\left(\frac{R}{a}\left[\frac{1}{\sqrt{1-\rho}}-1\right]\right) <1.
\end{align*} 
For the crust to be stable during the entire growth of a cavity, this inequality must hold for every $R$. Since $Q_0(x)$ decreases monotonically faster than $1/x$, it means that if the criterion is obeyed for $R=a$, it will also be obeyed for any $R>a$. Setting $L=a$ gives the following expression for the critical density
\begin{align}
\label{criterion_g0}
\rho_c= 1-\frac{1}{\left[1+Q_0^{-1}\left(\frac{1}{2 \pi }\right)\right]^2}\,,
\end{align}
where $Q_0(\cdot)^{-1}$ is the inverse function of $Q_0(\cdot)$.  Because $Q_0$ is unknown, it is necessary to resort to the numerical result given in Fig.~\ref{fig:2d_Crust_Analysis} to estimate the critical density.

Parenthetically, we can extend the above argument for the critical density to arbitrary spatial dimension $d$. The criterion for the crust thickness now is $\rho(R+z)^d= \left[(R+z)^d-R^d\right]$, which gives
$z=R\left[1/(1-\rho)^{1/d}-1\right]\equiv A_d R$. The average number of obstacles stacked at a given angular position along the crust is $\langle n\rangle = A_dR/a$, and the number of independent locations on the crust is $\mathcal{N}=\Omega_{d-1}(R/a)^{d-1}$, where $\Omega_d$ is the volume of a unit sphere in $d$ dimensions. Again under the assumption that crust growth is a random process, the $d$-dimensional version of the stability criterion \eqref{criterion_g0} is
\begin{align*}
\Omega_{d-1} \left(\frac{R}{a}\right)^{d-1}\;
Q_0\left(\frac{R}{a}\left[\frac{1}{(1-\rho)^{1/d}}-1\right]\right) <1.
\end{align*} 
For large spatial dimensions, the leading factor increases rapidly with $R/a$ while $Q_0$ should have a similar dependence on its argument as in two dimensions.  Thus the product may violate the above bound.  This would lead to instability of the crust unless the density is very close to 1.

\textit{Conclusion and outlook---} To summarize, we introduced the pushy random walk model to describe the motion of a mobile tracer particle that moves stochastically in a dense medium in which obstacles can be displaced due to impacts by the tracer particle. These impacts can lead to the formation of contiguous clusters of obstacles that become progressively harder to push as they grow. More importantly, these rearrangements of the background medium can extend over a macroscopic spatial range. The pushy random walk also extends the Sokoban walk model by replacing the sharp “single-obstacle” cutoff in the Sokoban walk with a size-dependent pushing rule that more closely mirrors the coupling between a tracer and the background medium that has been observed experimentally~\cite{Altshuler2024}.

In the Sokoban walk, the ability to move at most single obstacles ultimately guarantees self-caging: regardless of the initial density, the walker eventually encounters an unpushable configuration and becomes trapped. By contrast, the pushy random walk allows the tracer to displace clusters of any size, so that escape from local cages is possible through large rearrangements of obstacles. This softened constraint reshapes the familiar percolation scenario: without pushing, the square-lattice percolation threshold ($\rho_p\approx 0.407$) separates diffusion for obstacle densities $\rho<\rho_p$ from confinement for $\rho>\rho_p$.  When pushing can occur, the transition both moves and its character changes: shifting to a higher density ($\rho_c\approx 0.71$) and replacing confinement by subdiffusive spreading in the high-density phase, where the tracer is trapped in a cavity that continues to grow slowly in time.

We showed that the high-density phase is characterized by a slowly expanding cavity whose radius grows as $R(t)\sim t^{1/(3+\alpha)}$ (and $R(t)\sim t^{1/4}$ for $\alpha=1$), and that the transition is governed by a simple geometric criterion: the surrounding crust becomes effectively impenetrable when the expected number of holes along its circumference is less than $1$. In one dimension, by contrast, the pushy random walk always carves out an obstacle-free cavity whose length grows subdiffusively with time, $L(t)\sim t^{1/3}$ when the pushing power of the tracer varies inversely with the mass $M$ of the obstacle (and more generally, as $L(t)\sim t^{1/(2+\alpha)}$). Together, these results show how tracer-induced rearrangements qualitatively reshape transport in crowded media.

The situation in dimensions greater than two is not settled and merits more careful study. In particular, the $d$-dimensional criterion derived above should be viewed only as a formal extrapolation of the crust-stability argument, rather than a validated prediction for $d>2$. Detailed simulations will be required to assess whether the same diffusion-subdiffusion transition holds in higher dimensions.

\textit{Acknowledgments—} This project has received funding from the European Research Council (ERC) under the European Union’s Horizon 2020 research and innovation program (Grant Agreement No. 947731).

\textit{Data availability—} There are no publicly available research data or software supporting this manuscript. Requests for further information or data should be sent to the authors.

\end{document}